\documentclass[12pt,reqno]{amsart}
\usepackage{amscd,amssymb,graphics,color,a4wide,hyperref,verbatim}
\usepackage{graphicx}
\usepackage{psfrag} 
\usepackage{marvosym}

\usepackage{mathrsfs}
\input xy
\xyoption{all}

\def\Tr{\,{\rm Tr}\, }

\def\be{\begin{equation}}
\def\ee{\end{equation}}
\def\ba{\begin{eqnarray}}
\def\ea{\end{eqnarray}}

\renewcommand{\H}{{\mathcal H}}

\newcommand{\N}{{\mathcal N}}

\newcommand{\V}{\mathcal V}

\newcommand{\RR}{\mathbb{R}}

\newcommand{\ZZ}{\mathbb{Z}}

\newcommand{\CC}{\mathbb{C}}

\newcommand{\HH}{\mathbb{H}}

\DeclareMathOperator{\ch}{ch}

\newlength{\picwidth} 
\newlength{\miniwidth} 
\newlength{\nanowidth} 
\newlength{\melowidth} 
\newlength{\leftminiwidth} 
\newlength{\rightminiwidth} 
\newlength{\minipagewidth} 

\numberwithin{equation}{section}

\begin{document}

\title[Generalised Moonshine and Holomorphic Orbifolds]{
Generalised Moonshine and Holomorphic Orbifolds}
\thanks{{\it Prepared for the Proceedings of String Math 2012.}}

\author[]{Matthias R.~Gaberdiel}
\address{ Institut f\"ur Theoretische Physik, ETH Z\"urich, CH-8093 Z\"urich, Switzerland}
\email{gaberdiel@itp.phys.ethz.ch}

\author[]{Daniel Persson}
\address{Fundamental Physics, Chalmers University of Technology,
  412 96, Gothenburg, Sweden}
  \email{daniel.persson@chalmers.se}
  \urladdr{http://www.danper.se}

\author[]{Roberto Volpato}
\address{Max-Planck-Institut f\"ur Gravitationsphysik,
Am M\"uhlenberg 1, 14476 Golm, Germany}
\email{roberto.volpato@aei.mpg.de}

\date{2013-01-11}
\begin{abstract}
Generalised moonshine is reviewed from the point of view of holomorphic orbifolds, putting
special emphasis on the role of the third cohomology group  $H^3(G, U(1))$
in characterising consistent constructions. These ideas are then applied to the case
of Mathieu moonshine, i.e.\ the recently discovered connection between the largest Mathieu group 
$M_{24}$ and the elliptic genus of K3. In particular, we find a complete list of twisted twining genera
whose modular properties are controlled by a class in $H^3(M_{24}, U(1))$, as expected from 
general orbifold considerations.

\end{abstract}
\maketitle
\vspace{.2cm}
%
%
%
%
%
%
%
%
%
%
\section{Introduction}
Monstrous moonshine refers to a deep connection between modular forms, the Monster group $\mathbb{M}$, 
generalised Kac-Moody algebras and string theory. It
unfolded over the course of 15 years, starting with the Conway-Norton conjecture  in 1979 \cite{ConwayNorton} and 
the subsequent construction of the Frenkel-Lepowsky-Meurman Monster module   $\V^{\natural}$ \cite{FLM}, finally 
culminating in Borcherds complete proof of the moonshine conjecture \cite{Borcherds}. In a nutshell, monstrous moonshine 
asserts that for each element $g\in \mathbb{M}$ of the Monster group, there exists a class function $T_g$
(the McKay-Thompson series),  which is a holomorphic modular function (more precisely, the hauptmodul for a genus zero subgroup of $SL(2,\RR)$, see \cite{ConwayNorton}) on the upper-half-plane $\mathbb{H}$, and for which
the Fourier coefficients are characters of representations of $\mathbb{M}$. For example, 
when $g=e$ (the identity element), the McKay-Thompson series $T_e$  coincides with the modular-invariant $J$-function whose 
coefficients are dimensions of Monster group representations.

A few years after the original moonshine conjectures, Norton proposed \cite{Norton} an extension  that he dubbed generalised monstrous moonshine. 
Norton argued that to each commuting pair $(g,h)$ of elements in $\mathbb{M}$ there should exist a holomorphic modular function $f(g,h;\tau)$ 
on $\mathbb{H}$, whose Fourier coefficients also carry representation-theoretic information about the Monster. The generalised moonshine
conjecture was subsequently interpreted physically by Ginsparg, Dixon, Harvey \cite{Dixon:1988qd} in terms of orbifolds of the Monster CFT $\V^{\natural}$. Although the conjecture
has been proven for many special cases \cite{Tuite:1994ni,Dong:1997ea,Hohn}, the general case remains open (see however \cite{CarnahanI,CarnahanII,CarnahanIII,CarnahanIV} for recent progress).

A  new moonshine phenomenon was conjectured in 2010 by Eguchi, Ooguri, Tachikawa (EOT) \cite{Eguchi:2010ej}, 
subsequently dubbed Mathieu moonshine. In this case the Monster group $\mathbb{M}$ is replaced by the largest 
Mathieu group $M_{24}$, and the role of the modular $J$-function is played by the unique weak Jacobi form 
$\phi_{0,1}(\tau, z)$ of weight $0$ 
and index $1$ corresponding to the elliptic genus of K3. The analogue of the McKay-Thompson series, the so called 
twining genera $\phi_g(\tau, z)$, $g\in M_{24}$, were constructed in a series of papers 
\cite{Cheng:2010pq,Gaberdiel:2010ch,Gaberdiel:2010ca,Eguchi:2010fg}, and it was verified that they have precisely the 
properties required for Mathieu moonshine to hold. Indeed,  Gannon has recently shown \cite{GannonMathieu} that all 
multiplicity spaces can be consistently decomposed into sums of irreducible representations of $M_{24}$, 
thereby proving the EOT conjecture. 

Although  this establishes Mathieu moonshine, there is a major outstanding question: what is the $M_{24}$-analogue of the 
Monster module $\V^{\natural}$? 
In \cite{Gaberdiel:2012gf} we gave  evidence that some kind of holomorphic vertex operator algebra (VOA)
should be underlying Mathieu moonshine. The main point was to extend the previous results on twining genera to the 
complete set of twisted twining genera $\phi_{g,h}(\tau, z)$, corresponding to the $M_{24}$-analogues of Norton's generalised  moonshine 
functions $f(g,h;\tau)$ for the Monster. One of the key insights was that many of the properties of these functions, such 
as modularity, are controlled by a class in the third cohomology group $H^3(M_{24}, U(1))$, just as for orbifolds of 
holomorphic VOAs \cite{Dijkgraaf:1989pz,Roche:1990hs,Bantay:1990yr}.   

Our aim in this note is to give a short review of the generalised Mathieu moonshine phenomenon uncovered 
in \cite{Gaberdiel:2012gf}, focussing on the main ideas rather than technical details. For completeness we  
include a discussion of  holomorphic orbifolds and group cohomology which are the key ingredients in our work, 
as well as some background on Norton's generalised moonshine conjecture, which served as strong motivation for 
\cite{Gaberdiel:2012gf}.

This short note is organised as follows. We begin in section \ref{holorb} by discussing some features of orbifolds of  
holomorphic VOAs, explaining in particular the 
crucial role played by the cohomology group $H^3(G, U(1))$. 
In section \ref{GenMath} we then proceed to discuss generalised Mathieu moonshine. We define the twisted twining genera
and list the properties they should satisfy. We show that there is a unique class in $H^3(M_{24}, U(1))$ that is compatible
with the modular properties of the twining genera, and use this input to construct all twisted twining genera 
explicitly. Finally, we end in section \ref{concl} with a brief summary. 
\section{Holomorphic Orbifolds and Generalised Moonshine}
\label{holorb}
 In this section we will review some pertinent properties of orbifolds of 
holomorphic VOAs, with particular focus on the role of group cohomology. 

\subsection{Preliminaries}
Let $\V$ be a rational vertex operator algebra (VOA), and let $\mathcal{H}$ be a $\mathbb{Z}$-graded $\V$-module.\footnote{See for instance \cite{Terrybook} for a nice introduction to VOAs.} 
Rationality implies that $\V$ has only finitely many inequivalent simple modules $\mathcal{H}$, and that each graded 
component of $\mathcal{H}$ is finite-dimensional. By a \emph{holomorphic} (or `self-dual') VOA we shall mean the case 
that $\V$ has a \emph{unique} such module, namely the adjoint module of $\V$ itself; in this case we shall also write
$\V$ for this  module. The partition function of a holomorphic 
VOA is a holomorphic section of a line bundle over the moduli space of Riemann surfaces. The most prominent example of 
a holomorphic VOA is the moonshine module $\V^{\natural}$ \cite{FLM}, to which we shall return below.

Suppose we have a holomorphic VOA $\V$  with finite automorphism group $G$.
We want to analyse the orbifold of $\V$ by $G$, denoted $\widehat{\V}=\V/G$. The first step consists in projecting onto the 
$G$-invariant sub-VOA
\be
\V^G=\{\psi\in \V \ | \ g\psi=\psi, \forall g\in G\}\ .
\ee
The character of the VOA $\V^G$ 
is however not modular invariant, and to remedy this we must include twisted sectors. 
Since $\V$ is holomorphic the twisted sectors are just labelled by conjugacy classes in $G$, i.e.\ for each 
 $g\in G$ there is a $g$-twisted simple $\V$-module (or $g$-twisted sector) $\H_g$ \cite{Dong:1997ea}, 
which is an ordinary module for the $G$-invariant sub-VOA $\V^G$. The twisted sectors associated to group elements
in the same conjugacy class are isomorphic.

Each automorphism $h\in G$ of the VOA $\V$ induces a linear map $\H_g\to \H_{hgh^{-1}}$ between twisted sectors. 
In particular, each twisted sector $\H_g$ carries a representation of the centraliser
\be 
C_G(g):=\{ h\in G | hgh^{-1}=g\}\subseteq G
\ee 
 of $g$ in $G$,  though in general this will not be an honest representation.
 We will discuss this important subtlety below.

\subsection{Twisted Twining Characters}

Given a holomorphic VOA $\V$ of central charge $c$ its partition function is defined by the usual formula
\be
\mathscr{Z}_\V(\tau)=\text{Tr}_{\V} (q^{L_0-c/24}) \ , 
\ee
where $q=e^{2\pi i \tau}$ and  $L_0$  is the Virasoro (Cartan) generator. Similarly, for each twisted sector 
$\H_g$  in the orbifold theory one may construct the associated twisted character (sometimes called `characteristic function')
\be
Z_{g, e}(\tau)=\text{Tr}_{\H_g}(q^{L_0-c/24}) \ , 
\ee
where $e$ denotes the identity element in $G$. Moreover, since $\H_g$ is invariant under the centraliser subgroup 
$C_G(g)$ it makes sense to define, for all $h\in C_G(g)$, the twisted twining character
\be
Z_{g,h}(\tau)=\text{Tr}_{\H_g}\big(\rho(h) \, q^{L_0-c/24}\big) \ ,
\ee
where $\rho : C_G(g)\to \text{End}(\H_g)$ denotes the representation with which $h$  acts on the twisted vector space $\H_g$. 

Physically, the twisted twining character $Z_{g,h}$ corresponds to the path integral on a torus  with modular parameter $\tau$ 
and boundary conditions twisted by $(g,h)$ along the $(a,b)$-cycles of $\mathbb{T}^2$. Choosing periodic boundary conditions 
corresponds to setting $(g,h)=(e,e)$ and hence gives back the original partition function
\be
Z_{e,e}(\tau)=\mathscr{Z}(\tau)\ .
\ee 
%
Given the definition of the twisted twining characters $Z_{g,h}$ one should expect that
they only depend on the 
conjugacy class of $(g,h)$ in $G$, i.e.\ they should correspond to class functions
\be
Z_{g,h}(\tau)=Z_{k^{-1}gk, k^{-1}hk}(\tau)\ , \qquad \qquad k\in G\ .
\label{conj}
\ee
As we shall see (see section \ref{twisted} below), in general this property will only be true up to
a phase.

In contrast to $\mathscr{Z}(\tau)$, the twisted twining characters $Z_{g,h}(\tau)$ are not invariant under the full modular group $SL(2,\mathbb{Z})$. Under a modular transformation 
\be
\tau \longmapsto \frac{a\tau+b}{c\tau+d}\ , \qquad \qquad \left(\begin{array}{cc} a & b \\ c & d \\ \end{array} \right)
\in SL(2,\mathbb{Z})\ , 
\ee
the spin structures of the torus change such that the twists by $g$ and $h$  along the $a$- and $b$-cycles transform according to
\be
(g,h) \longmapsto (g,h)\left(\begin{array}{cc} a & b \\ c & d \\ \end{array} \right)^{-1}=(g^d h^{-c}, g^{-b} h^a)\ .
\ee
The  twisted twining characters then transform among themselves as 
\be
Z_{g,h}\left(\frac{a\tau+b}{c\tau+d}\right)=\chi_{g,h}(\begin{smallmatrix} a & b \\ c & d  \end{smallmatrix}) \,
Z_{g^a h^{c}, g^{b} h^d}(\tau)\ ,
\label{modularprop}
\ee
where we have included the possibility of having a non-trivial multiplier system 
\be
\chi_{g,h}\, :\, SL(2,\mathbb{Z})\, \longrightarrow \, U(1)\ .
\ee
The set of functions $\{Z_{g,h}\}$ thus forms a representation of $SL(2, \mathbb{Z})$.

\subsection{Twisted Sectors and Projective Representations}
\label{twisted}
As we have mentioned above, the states in the twisted sector $\H_g$ transform in a representation 
$\rho$ of $C_G(g)$. However, 
this representation need not be 
an honest representation, but may only be projective. Recall that a projective representation
$\rho$ of a finite group $H$  respects the group multiplication only up to a phase,
\be
\rho(h_1) \, \rho(h_2)=c(h_1, h_2)\, \rho(h_1 h_2)\ ,
\ee
where $c(h_1,h_2)$ is a $U(1)$-valued 2-cocycle, representing a class in $H^2(H, U(1))$. Thus we have,
for each twisted sector $\H_g$, a class $c_g\in H^2(C_G(g),U(1))$, characterising the projectivity of the 
action of $C_G(g)$ in the 
$g$-twisted sector. One consequence of these  phases is that 
the  formula (\ref{conj}) must be modified; the correct generalisation is 
 \be
Z_{g, h}(\tau) = \frac{c_g(h,k)}{c_g(k, k^{-1}h k)} \, Z_{k^{-1}gk, k^{-1}hk}(\tau)\ .
\label{conj2}
\ee
Furthermore, these phases modify the modular $S$ and $T$-transformations as  \cite{Bantay:1990yr}
\be
\begin{array}{rcl}
{Z}_{g,h}(\tau+1)&=& c_g(g,h)\, {Z}_{g, gh}(\tau)\ ,\\[4pt]
{Z}_{g,h}(-1/\tau)&=& \overline{c_h(g, g^{-1})}\, {Z}_{h, g^{-1}}(\tau)\ .
\end{array}
\label{STmodular}
\ee
Although the twisted twining genera $Z_{g,h}$ are not invariant under the full $SL(2,\mathbb{Z})$, they will be 
modular functions with respect to 
some arithmetic subgroup $\Gamma_{g,h}\subset SL(2,\mathbb{Z})$ which fixes the pair $(g,h)$. The 
group $\Gamma_{g,h}$ will, in particular, contain 
a congruence subgroup $\Gamma(N)\subset SL(2,\mathbb{Z})$, for a suitable positive integer $N$. 
Each $Z_{g,h}$ therefore has a Fourier expansion of the form
\be
Z_{g,h}(\tau)=\sum_{n} \text{Tr}_{\H_{g,n}}\left(\rho(h)\right) q^{n/N},
\ee
where $\H_{g,n}$ is the grade $n$ subspace of the twisted module $\H_g$ and $\text{Tr}_{\mathcal{H}_{g,n}}\left(\rho(h)\right)$  
is a projective character of $C_G(g)$, i.e.\ a character of a graded representation of a central extension of $C_G(g)$.
We can
therefore decompose the different graded components 
$\H_{g,n}$ of the twisted module $\H_g$ into (finite) sums of irreducible projective representations $R_j$,
each corresponding to the same $2$-cocycle class $c_g$
\be
\H_{g, n}=\bigoplus_{j}h_{g,n}^{(j)} R_j\ .
\label{decproj}
\ee
Here $h_{g,n}^{(j)}$ describes the multiplicity with which $R_j$  occurs.

\subsection{Group Cohomology of Holomorphic Orbifolds}
In the previous section we have seen that the action of $h\in G$ on the twisted sector $\H_g$ is generically projective, 
and inequivalent choices are classified by $H^2(C_G(g), U(1))$. The appearance of this cohomology group can in fact be 
traced back to an even finer and more sophisticated underlying structure, namely the third cohomology group $H^3(G, U(1))$. 

For every commuting pair $g,h\in G$ the fusion product between the associated twisted sectors induces an 
isomorphism $\H_g \boxtimes \H_h \rightarrow \H_{gh}$. For every triple $g,h,k\in G$ there exists a 
3-cocycle $\alpha(g,h,k)\in H^3(G, U(1))$ which measures the failure of associativity in the choice of isomorphism 
for the triple fusion product \cite{Roche:1990hs}. The third cohomology group  therefore classifies consistent 
holomorphic orbifolds. The class $[\alpha] \in H^3(G, U(1))$ determines many properties of the 
orbifold theory $\widehat{\V}$. In particular, it determines the particular central extension of $C_G(g)$  which 
controls the projective representations $\rho$ in the $g$-twisted sector $\H_g$. Indeed, for 
every $h\in G$, the 3-cocycle $\alpha$ gives rise to a distinguished element $c_h\in H^2(C_G(h), U(1))$ through 
the formula  \cite{Dijkgraaf:1989pz,Bantay:1990yr}
\be 
c_h(g_1, g_2)= \frac{\alpha(h, g_1, g_2)\,\, \alpha(g_1, g_2,(g_1g_2)^{-1} h(g_1g_2))}{\alpha(g_1, h, h^{-1}g_2h)}\ .
\label{ch}
\ee 
Since the projective phases also control the modular properties of $Z_{g,h}$ (\ref{STmodular}), these are 
then also determined in terms of the class $[\alpha]$. 

These phases actually lead to a number of interesting consequences. For example, for the special case when
$k, g$ and $h$ are pairwise commuting elements in $G$, we get 
\be
Z_{g,h}(\tau)= \frac{c_g(h,k)}{c_g(k, h )} \, Z_{g, h}(\tau) \ ,
\ee
and thus $Z_{g,h}(\tau)=0$ unless the $2$-cocycle satisfies the regularity condition
\be\label{regularity}
c_g(h,k)=c_g(k,h) 
\ee
for all $k\in G$ that commute both with $g$ and $h$. 


\subsection{Application to Generalised Monstrous Moonshine}
We will now discuss a specific example of the framework introduced above, that arises for the case when 
$\mathcal{V}$ is the Frenkel-Lepowsky-Meurman (FLM) Monster VOA  $\V^{\natural}$ \cite{FLM} with $c=24$, whose 
automorphism group $G$ is the Monster group $\mathbb{M}$.

In 1987 Norton proposed \cite{Norton} a generalisation of monstrous moonshine in which he suggested that it was natural to 
associate a holomorphic function $f(g,h;\tau)$ to each commuting pair $(g,h)$ of elements in $\mathbb{M}$. Norton 
argued that these functions should satisfy the following conditions:
\begin{enumerate}
\item[(1)] $f(g,h;\tau)=f(k^{-1}gk, k^{-1}hk;\tau)\ , \qquad \qquad k\in \mathbb{M}$
\item[(2)] $ f(g,h;\tau)=\gamma f\big(g^{a}h^{c}, g^b h^d;\tfrac{a\tau+b}{c\tau+d}\big)\ , \qquad  (\begin{smallmatrix} a & b \\ c & d  \end{smallmatrix})\in SL(2,\mathbb{Z})$\\
{\it (Here $\gamma$ is a 24'th root of unity.)}
\item[(3)] {\it the coefficients in the $q$-expansion of $f(g,h;\tau)$ are characters of a graded projective representation of $C_\mathbb{M}(g)$}
\item[(4)] {\it $f(g,h;\tau)$ is either constant or a hauptmodul for some genus zero $\Gamma_{g,h}\subset SL(2,\mathbb{R})$}
\item[(5)] {\it $f(e,h;\tau)=T_h(\tau)$, where $T_h$ is the McKay-Thompson series associated to $h$}
\end{enumerate}

All of these conditions, with the exception of the genus zero property $(4)$, can be understood within the 
framework of holomorphic orbifolds. The FLM Monster module $\V^{\natural}$ is a holomorphic VOA and so 
for each $g\in \mathbb{M}$ we have a unique $g$-twisted module $\H_g^{\natural}$ with an inherited grading 
\be
\H_g^{\natural}=\bigoplus_{n=-N}^{\infty} \H_{g,n}^{\natural}\ ,
\ee
where each $\H_{g,n}^{\natural}$ is a projective representation of $C_{\mathbb{M}}(g)$. 
For each twisted module $\H_g^{\natural}$ we can define the associated twisted twining character
\be
Z^{(\V^{\natural})}_{g,h}(\tau)=\text{Tr}_{\H_{g}^{\natural}}( \rho(h) \, q^{L_0-1}) 
=\sum_{n=-N}^{\infty} \text{Tr}_{\H^{\natural}_{g,n}}\big(\rho(h)\big)\, q^{n/N},
\ee
for a suitable positive integer $N$. 
By the  properties of holomorphic orbifolds discussed above, this twisted twining character 
satisfies properties $(1)-(3)$ and $(5)$ of Norton, and 
it is therefore natural to suspect that $Z^{(\V^{\natural})}_{g,h}(\tau)=f(g,h;\tau)$. This 
connection was first made by Dixon, Ginsparg and Harvey \cite{Dixon:1988qd}, and  has subsequently 
been proven in many special cases \cite{Tuite:1994ni,Dong:1997ea,Hohn}, though the general conjecture 
remains open.

Since generalised moonshine can be understood within the framework of holomorphic orbifolds, one should expect that
the third  cohomology group $H^3(\mathbb{M}, U(1))$ plays an important role (see \cite{MasonCohomology} for a related discussion). In particular, one might guess 
that Norton's condition $(1)$ should be generalised to include the cohomological prefactor (\ref{conj2}), 
involving a 2-cocycle $c_g\in H^2(C_{\mathbb{M}}(g), U(1))$. Moreover, the roots of unity $\gamma$ appearing 
in the modular transformation $(2)$ should be computable from some $\alpha\in H^3(\mathbb{M}, U(1))$, via 
the general formulae (\ref{STmodular}). This would then also suggest that the cases where $f(g,h;\tau)$ are constant
(see condition $(4)$) are manifestations of an obstruction, for example of the type described above in 
(\ref{regularity}).\footnote{We thank Terry Gannon for suggesting this idea to us, see also
\cite{GannonMathieu}.} Unfortunately, little is known 
about $H^3(\mathbb{M}, U(1))$, and thus it is difficult to confirm this directly.


\vspace{.5cm} 

\section{Generalised Mathieu Moonshine}
\label{GenMath}
\subsection{A Lightning Review of Mathieu Moonshine}

In 2010, Eguchi, Ooguri and Tachikawa \cite{Eguchi:2010ej} conjectured a supersymmetric version of the 
moonshine phenomenon for a certain sporadic finite simple group, the Mathieu group $M_{24}$, where the role 
of the  $J$-function is played by the elliptic genus of K3. The latter is most naturally defined as a refined partition 
function of a certain class of two-dimensional superconformal field theories with $\N=(4,4)$ superconformal symmetry, which 
have central charge $c=6$, and can be realised as non-linear sigma models with target space 
K3. More precisely, the elliptic genus is a complex function on $\HH\times \CC$ defined as
\be 
\phi_{\rm K3}(\tau,z)=\Tr_{RR}\bigl((-1)^{F+\tilde F}q^{L_0-\frac{c}{24}}\bar{q}^{\tilde{L}_0-\frac{\tilde c}{24}}
y^{2J_0^3}\bigr)\ ,\qquad q=e^{2\pi i\tau},\quad y=e^{2\pi i z}\ ,
\ee
where $L_0,\tilde{L}_0$ are the left- and right-moving Virasoro generators, $(-1)^{F+\tilde{F}}$ is the total worldsheet fermion number, 
$J_0^3$ is the 
Cartan generator of the affine $su(2)_1$ subalgebra of the left $\N=4$ superconformal algebra, and
the trace is taken over the Ramond-Ramond sector $\H_{\rm RR}$ of the theory. In general, the elliptic genus can be defined in any 
theory with (at least) $\N=2$ superconformal symmetry and does not change under superconformal deformations of the theory. In a 
non-linear sigma model, this means that $\phi$ is independent of the choice of a metric and the Kalb-Ramond field of the target 
space, but it encodes information on the topology. For example, $\phi(\tau,z=0)$ is the Euler number of the target space, so that 
in particular $\phi_{\rm K3}(\tau,0)=24$. 
 
The only states that give rise to a non-vanishing contribution to $\phi_{\rm K3}$ are the right-moving ground states, i.e.\ the
eigenstates with zero eigenvalue for $\tilde{L}_0-\frac{1}{4}$; this implies that K3 is holomorphic both in $\tau$ and $z$. 
The elliptic genus has good modular properties
\be 
\phi_{g,h}\Bigl(\frac{a \tau + b}{c \tau + d} , \frac{z}{c \tau + d}\Bigr) =
e^{ 2 \pi i  \frac{c z^2}{c \tau + d} } \, 
\phi_{\rm K3}(\tau,z) \ ,\qquad (\begin{smallmatrix} a & b \\ c & d  \end{smallmatrix} )\in SL(2,\ZZ) \ ,
\ee 
and because of the spectral flow automorphism of the $\N=4$ superconformal algebra, possesses the elliptic 
transformation rules \cite{EOTY}
\be 
\phi_{\rm K3}(\tau,z+ \ell \tau + \ell') = e^{-2 \pi i(\ell^2 \tau+ 2 \ell z)}\, \phi_{\rm K3}(\tau,z) \qquad 
\ell,\ell'\in \ZZ\ .
\ee 
These are the defining properties of a weak Jacobi form of weight $0$ and index $1$ \cite{EichlerZagier}, and are 
sufficient to determine $\phi_{\rm K3}$ up to normalisation, which in turn is fixed by the condition $\phi_{\rm K3}(\tau,0)=24$. 
Explicitly,
\be \phi_{\rm K3}(\tau,z)=8\Bigl(\frac{\vartheta_2(\tau,z)^2}{\vartheta_2(\tau,0)^2}+
\frac{\vartheta_3(\tau,z)^2}{\vartheta_3(\tau,0)^2}+\frac{\vartheta_4(\tau,z)^2}{\vartheta_4(\tau,0)^2}\Bigr)\ ,
\ee in terms of Jacobi theta functions \cite{EichlerZagier}.
The states contributing to the elliptic genus form a representation of the left $\N=4$ superconformal algebra, so that 
$\phi_{\rm K3}$ admits a decomposition into irreducible $\N=4$ characters
\begin{align*}
\phi_{\rm K3}(\tau,z)=
&20 \ch_{\frac{1}{4},0}(\tau,z)-2\ch_{\frac{1}{4},\frac{1}{2}}(\tau,z)+\sum_{n=1}^\infty A_n \ch_{\frac{1}{4}+n,\frac{1}{2}}(\tau,z)\ .
\end{align*} Here,
$\ch_{h,\ell}(\tau,z)=\Tr_{h,\ell}((-1)^{F}q^{L_0-\frac{c}{24}}y^{2J_0^3})
$ is the character of the Ramond $\N=4$ representation whose highest weight vector is an eigenstate with 
eigenvalues $h,\ell$ under $L_0$ and $J_0^3$, respectively. By unitarity, the only possible values for 
$(h,\ell)$ are $(\frac{1}{4},0)$, $(\frac{1}{4},\frac{1}{2})$ (short or BPS representations), and 
$(\frac{1}{4}+n,\frac{1}{2})$, $n=1,2,3,\ldots$ \cite{Eguchi:1987sm,Eguchi:1987wf}. 
Finally, if $N(h,\ell;\bar h,\bar\ell)$ is the multiplicity of the 
corresponding $\N=(4,4)$ representation in the spectrum of the theory, then 
\be 
A_n:=\sum_{(h=\frac{1}{4}+n,\ell=\frac{1}{2};\bar h,\bar \ell)} \ch_{\bar h,\bar\ell}(\bar\tau,0)
=N(\tfrac{1}{4}+n,\tfrac{1}{2};\tfrac{1}{4},0)-2N(\tfrac{1}{4}+n,\tfrac{1}{2};\tfrac{1}{4},\tfrac{1}{2})
\ee 
is the $\ZZ_2$-graded multiplicity of the $(h,\ell)$ representations of the left $\N=4$ algebra. 
As it turns out, the $A_n$ with $n\geq 1$ are all even positive integers. The most surprising property, 
however, is that the first few of them
\be 
\tfrac{1}{2}A_n=45,\ 231,\ 770,\ 2277,\ 5796,\ \ldots
\ee 
exactly match the dimensions of some irreducible representations of $M_{24}$ \cite{Eguchi:2010ej}. In 
analogy with the monstrous moonshine observation, it is then natural to conjecture that the space of states 
contributing to the elliptic genus carries an action of $M_{24}$, commuting with the $\N=4$ algebra, 
so that 
\be 
\phi_{\rm K3}(\tau,z)=\sum_{(h,\ell)} \dim R_{h,\ell} \ch_{h,\ell}(\tau,z)\ ,
\ee 
for some (possibly virtual) $M_{24}$ representations $R_{h,\ell}$. Soon after the EOT observation, 
the analogues of the McKay-Thompson series, the twining genera
\be\label{twini} 
\phi_{g}(\tau,z)=\sum_{(h,\ell)} \Tr_{R_{h,\ell}}(g) \ch_{h,\ell}(\tau,z)\ ,\qquad g\in M_{24}
\ee  
have been considered \cite{Cheng:2010pq,Gaberdiel:2010ch}. Each $\phi_g$ is expected to be a Jacobi form of 
weight $0$ and index $1$ (possibly up to a multiplier \cite{Gaberdiel:2010ch,Gaberdiel:2010ca}) for a group
\be 
\Gamma_0(N):=\left\{\begin{pmatrix} a & b\\ c & d
\end{pmatrix}\in SL(2,\ZZ)\mid c\equiv 0\mod N \right\}\ ,
\ee
where $N=o(g)$ is the order of $g$. Explicitly \cite{Gaberdiel:2010ca},
\begin{align}\label{modtwin1}
&\phi_{g}(\tau,z+ \ell \tau + \ell') = e^{-2 \pi i(\ell^2 \tau+ 2 \ell z)}\, \phi_{g}(\tau,z) &
\ell,\ell'\in \ZZ\\
&\phi_{g}\Bigl(\frac{a \tau + b}{c \tau + d} , \frac{z}{c \tau + d}\Bigr) =
e^{2\pi i\frac{cd}{N\ell(g)}}\,
e^{ 2 \pi i  \frac{c z^2}{c \tau + d} } \, \phi_{g}(\tau,z) \ ,
& \hspace*{-0.9cm} (\begin{smallmatrix} a & b \\ c & d  \end{smallmatrix} )\in \Gamma_0(N)\ , 
\label{modtwin2}
\end{align} 
where $\ell(g)$ is the length of the shortest cycle of $g$ in the $24$-dimensional permutation representation of 
$M_{24}$ \cite{Cheng:2011ay}. A complete list of twining genera satisfying \eqref{modtwin1} and \eqref{modtwin2} 
has been proposed in \cite{Gaberdiel:2010ca,Eguchi:2010fg}, where the first few hundred $M_{24}$-representation 
$R_{h,\ell}$ have been computed explicitly. Finally, it was shown in \cite{GannonMathieu} that all representations 
$R_{h,\ell}$ matching \eqref{twini} for all $g\in M_{24}$ exist, and that the the only virtual representations 
correspond to the BPS characters
\be\label{BPSreps} 
R_{\frac{1}{4},0}={\bf 23}-3\cdot {\bf 1}\ ,\qquad R_{\frac{1}{4},\frac{1}{2}}=-2\cdot {\bf 1}\ .
\ee 

These results, in a sense, prove the EOT conjecture. The interpretation of this Mathieu moonshine, however, 
is still an open problem. The most obvious explanation would be the existence of a non-linear sigma model on 
K3 with symmetry group $M_{24}$. If such a theory existed, the twining genera could be identified with the traces 
\be\label{twintrace} \phi_g(\tau,z)=\Tr_{RR}\bigl(g(-1)^{F+\tilde F}q^{L_0-\frac{c}{24}}
\bar{q}^{\tilde{L}_0-\frac{\tilde c}{24}}y^{2J_0^3}\bigr)\ ,
\ee 
and \eqref{modtwin1} and \eqref{modtwin2} would follow by standard CFT arguments. This possibility, however, has 
been excluded in \cite{Gaberdiel:2011fg}, where the actual groups of symmetries of non-linear sigma models on K3 
have been classified, and it was shown that none of them contains the Mathieu group $M_{24}$. More generally, one might 
conjecture the existence of some unknown CFT with $\N=(4,4)$ superconformal symmetry and carrying an action of 
$M_{24}$ such that the twining 
genera $\phi_g$ are reproduced by (\ref{twintrace}) 
for all $g\in M_{24}$. However, because of the $-3\cdot {\bf 1}$ in \eqref{BPSreps}, this theory should contain fields in the 
R-R representation $(h,\ell;\bar h,\bar\ell)=(\frac{1}{4},0;\frac{1}{4},\frac{1}{2})$ of the $\N=(4,4)$ algebra, which, 
by spectral flow, correspond to fields in the NS-NS representation $(h,\ell;\bar h,\bar\ell)=(\frac{1}{2},\frac{1}{2};0,0)$. 
It has been argued in \cite{NahmWend} that every theory with $\N=(4,4)$ superconformal symmetry at $c=6$ containing 
such fields is necessarily a non-linear sigma model on a torus, for which the elliptic genus vanishes. Thus it seems 
that a satisfactory explanation of Mathieu moonshine will need some more radically new idea.

\subsection{Twisted Twining Genera: Definitions and Properties}

As explained in the previous subsection, the twining genera $\phi_g$, $g\in M_{24}$ satisfy all the properties 
expected for traces of the form \eqref{twintrace} in a $\N=(4,4)$ theory with symmetry $M_{24}$. In such a (conjectural) 
theory, the twisted sector $\H_g$, for each $g\in M_{24}$, would form a representation $\rho_g$ of the centraliser 
$C_{M_{24}}(g)$, whose action commutes with the $\N=(4,4)$ superconformal algebra. The characters
\be\label{eqn:twining-trace}
\phi_{g,h}(\tau,z)=\Tr_{\H_g}\bigl(\rho_g(h)\,(-1)^{F+\tilde F}q^{L_0-\frac{c}{24}}\bar{q}^{\tilde{L}_0-\frac{\tilde c}{24}}y^{2J_0^3}\bigr)\ ,
\qquad g,h\in M_{24},\ gh=hg,
\ee 
would be the $\N=4$ counterpart of the twisted twining partition functions $Z_{g,h}$ considered in section 2, 
and should obey analogous properties.

As we have stressed above, a superconformal field theory with the properties above is not known. However, following the 
philosophy of the previous subsection, we will show that functions $\phi_{g,h}$ exist, satisfying all the properties expected 
for characters of the form \eqref{eqn:twining-trace}. This is very convincing evidence in favour of a generalised Mathieu 
moonshine, analogous to Norton's conjecture in the Monster case.

\bigskip

The definition of the twisted twining genera $\phi_{g,h}$ in terms of (\ref{eqn:twining-trace}) suggests that they
should satisfy the following properties:
\begin{enumerate}
\item Elliptic and modular properties:
\begin{align} 
&\phi_{g,h}(\tau,z+ \ell \tau + \ell') = e^{-2 \pi i(\ell^2 \tau+ 2 \ell z)}\, \phi_{g,h}(\tau,z) &
\ell,\ell'\in \ZZ\\
&\phi_{g,h}\Bigl(\frac{a \tau + b}{c \tau + d} , \frac{z}{c \tau + d}\Bigr) =
\chi_{g,h}(\begin{smallmatrix} a & b \\ c & d  \end{smallmatrix})\,
e^{ 2 \pi i  \frac{c z^2}{c \tau + d} } \, \phi_{g^ah^c,g^bh^d}(\tau,z) \ ,
& \hspace*{-0.9cm} (\begin{smallmatrix} a & b \\ c & d  \end{smallmatrix} )\in SL(2,\ZZ) 
\label{modcond}
\end{align} 
for a certain multiplier $\chi_{g,h}:SL(2,\ZZ)\to  U(1)$.
In particular, each $\phi_{g,h}$ is a weak Jacobi form of weight 0 and index 1 with multiplier $\chi_{g,h}$ 
under a subgroup $\Gamma_{g,h}$ of $SL(2,\ZZ)$.
\item Invariance  under conjugation of the pair $g,h$ in $M_{24}$, 
\be \label{conj1} 
\phi_{g,h}(\tau,z)=\xi_{g,h}(k)\, \phi_{k^{-1}gk,k^{-1}hk}(\tau,z)\ ,\qquad
k\in M_{24}\ ,
\ee
where $\xi_{g,h}(k)$ is a phase.
\item If $g\in M_{24}$ has order $N$, the twisted twining genera $\phi_{g,h}$ have an expansion of the form 
\be
\phi_{g,h}(\tau, z) = \sum_{\substack{r\in\lambda_g+\ZZ/N\\ r\ge 0}} \Tr_{\H_{g,r}}\bigl(\rho_{g,r}(h)\bigr)
\ch_{h=\frac{1}{4}+r, \ell}(\tau, z)\ , \label{eqn:decomp}
\ee 
where $\lambda_g\in\mathbb{Q}$, and $\ch_{h, \ell}(\tau,z)$ are elliptic genera of
Ramond representations of the $\N = 4$ superconformal algebra at central charge $c=6$. (Here $\ell=\tfrac{1}{2}$,
except possibly for $h=\tfrac{1}{4}$, where $\ell=0$ is also possible --- if both $\ell=0,\tfrac{1}{2}$ 
appear for $r=0$, it is understood that there are two such terms in the above sum.) Furthermore, each vector 
space $\H_{g,r}$ is finite dimensional, and it carries a projective representation $\rho_{g,r}$ of the centraliser 
$C_{M_{24}}(g)$ of $g$ in $M_{24}$, such that 
\begin{equation}
\rho_{g,r}(g) = e^{2\pi i r} \ , \qquad 
\rho_{g,r}(h_1) \, \rho_{g,r}(h_2) = c_g(h_1,h_2) \, \rho_{g,r}(h_1 h_2) \ ,
\end{equation}
for all $h_1,h_2\in C_{M_{24}}(g)$. Here $c_{g}: C_{M_{24}}(g) \times C_{M_{24}}(g) \to U(1)$ 
is independent of $r$, and satisfies the cocycle condition
\be 
c_g(h_1,h_2)\, c_g(h_1h_2,h_3)=c_g(h_1,h_2h_3)\, c_g(h_2,h_3)
\ee
for all $h_1,h_2,h_3\in C_{M_{24}}(g)$.
\item For $g=e$, where $e$ is the identity element of $M_{24}$, the functions $\phi_{e,h}$ correspond to the 
twining genera \eqref{twini}. In particular, $\phi_{e,e}$ is the K3 elliptic genus. 
\item The multipliers $\chi_{g,h}$, the phases $\xi_{g,h}$, and the 2-cocycles 
$c_g$ associated with the projective representations $\rho_{g,r}$ are completely determined 
(by the same formulas as for holomorphic orbifolds) in terms of a 
$3$-cocycle $\alpha$ representing a class in $H^3(M_{24},U(1))$.
\end{enumerate}

\subsection{The role of $H^3(M_{24}, U(1))$, obstructions and computation of $\phi_{g,h}$}\label{s:obst}

The third cohomology group of $M_{24}$ was only recently computed with the result  \cite{DutourEllis}\footnote{Note that 
for a finite group $G$ one has the isomorphisms 
$$H_{n-1}(G,\mathbb{Z})\cong H^n(G,\mathbb{Z}), \qquad \qquad H^{n}(G,\mathbb{Z})\cong H^{n-1}(G,U(1))\ ,$$
which in particular imply that $H_3(M_{24}, \mathbb{Z})\cong H^3(M_{24}, U(1)).$}
\be\label{cohomo}
H^3(M_{24}, U(1))\cong \ZZ_{12}\ .
\ee
The fact that this group is known explicitly plays a crucial role in our analysis. The specific cohomology 
class $[\alpha] \in H^3(M_{24},U(1))$ that is relevant in our context is uniquely determined by the condition 
that it reproduces the multiplier system for the 
twining genera $\phi_{e,h}$ as described in \cite{Gaberdiel:2010ca}, namely
\be \label{multiplier}
\chi_{e,h}(\begin{smallmatrix} a & b \\ c & d  \end{smallmatrix})=e^{\frac{2\pi i cd}{o(h)\ell(h)}}\ ,\qquad 
\left(\begin{smallmatrix} a & b \\ c & d  \end{smallmatrix}\right)\in \Gamma_0(o(h))\ .
\ee 
Here, $o(h)$ is the order of $h$ and $\ell(h)$ is the length of the smallest cycle, when $h\in M_{24}$ is regarded 
as a permutation of $24$ symbols \cite{Cheng:2011ay}.
Indeed, since $\ell({\rm 12B})=12$, it follows that 
$\alpha$ must correspond to a generator of $H^3(M_{24}, U(1))$. With the help of the software GAP \cite{GAP4}, 
we have verified that a generator reproducing the multiplier phases (\ref{multiplier})
exists and is unique \cite{Gaberdiel:2012gf}.

Once the $3$-cocycle $\alpha$ is known, one can use \eqref{modcond} and \eqref{conj1} to deduce the precise 
modular properties of each twisted twining genus $\phi_{g,h}$. It turns out that, in many cases, these properties can 
only be satisfied if $\phi_{g,h}$ vanishes identically \cite{Gaberdiel:2012gf}. In particular, there are two kinds of 
potential obstructions that can force a certain twisted twining genus to vanish
\begin{list}{(\roman{enumi})}{\usecounter{enumi}}
\item Consider three pairwise commuting elements $g,h,k\in M_{24}$. By \eqref{conj1},
\be \phi_{g,h}(\tau,z)=\xi_{g,h}(k) \phi_{g,h}(\tau,z)\ .
\ee 
Therefore, if $\xi_{g,h}(k)\neq 1$, we conclude that $\phi_{g,h}(\tau,z)=0$.
\item
Consider a commuting pair of elements $g,h\in M_{24}$, and suppose that $k\in M_{24}$ exists such that 
$k^{-1}g^{-1}k=g$ and $k^{-1}h^{-1}k=h$. Then, by \eqref{conj1} and \eqref{modcond}
\be 
\phi_{g,h}(\tau,-z)=\chi_{g,h}\bigl(\begin{smallmatrix} -1 & 0\\ 0 & -1
\end{smallmatrix}\bigr)\, \phi_{g^{-1},h^{-1}}(\tau,z)
=\chi_{g,h}\bigl(\begin{smallmatrix} -1 & 0\\ 0 & -1
\end{smallmatrix}\bigr)\, \xi_{g^{-1},h^{-1}}(k)\, \phi_{g,h}(\tau,z)\ .
\ee 
By \eqref{eqn:decomp}, and using the fact that the $\N=4$ characters are even functions of $z$, i.e.\ 
$\ch_{h,\ell}(\tau,-z)=\ch_{h,\ell}(\tau,z)$, we obtain 
\be\label{even}\phi_{g,h}(\tau,-z)=\phi_{g,h}(\tau,z)\ .
\ee 
Therefore, if 
\be 
\chi_{g,h}\bigl(\begin{smallmatrix} -1 & 0\\ 0 & -1
\end{smallmatrix}\bigr)\, \xi_{g^{-1},h^{-1}}(k)\neq 1\ ,
\ee 
$\phi_{g,h}$ must vanish.
\end{list}

\medskip

\noindent 
In all cases where $\phi_{g,h}$ is not obstructed, we define $\Gamma_{g,h}\subseteq SL(2,\ZZ)$ to be the 
subgroup of $SL(2,\ZZ)$ that leaves $(g,h)$ fixed or maps it to $(g^{-1},h^{-1})$, up to conjugation in $M_{24}$, i.e.\
\be 
\Gamma_{g,h}=\{(\begin{smallmatrix} a & b \\ c & d  \end{smallmatrix})\in 
SL(2,\ZZ)\mid\exists k\in M_{24}, (g^ah^c,g^bh^d)=(k^{-1}gk,k^{-1}hk)\text{ or }(k^{-1}g^{-1}k,k^{-1}h^{-1}k)\}\ .
\ee 
Then, by \eqref{modcond}, \eqref{conj1} and \eqref{even}, $\phi_{g,h}$ must be a weak Jacobi form of weight 
$0$ and index $1$ under $\Gamma_{g,h}$, possibly with a multiplier. It turns out that, whenever $g\neq e$, the 
spaces of such Jacobi forms are either zero- or one-dimensional, and the normalisation can be easily fixed by 
requiring that a decomposition of the form \eqref{eqn:decomp} exists (note that, since the representations 
$\rho_{g,r}$ are projective, the phase of the normalisation is ambiguous). This allows us to determine 
$\phi_{g,h}$ for all commuting pairs $g,h\in M_{24}$. The results are summarised in the next subsection.

\subsection{Generalised Mathieu Moonshine: Statement of Results}

In order to describe the twisted twining genera $\phi_{g,h}$ for all commuting pairs of elements $g,h\in M_{24}$, we
first note that the functions associated to different such pairs are not necessarily independent. 
In particular, because of 
 \eqref{modcond} and \eqref{conj1}, we have relations between pairs that are conjugated by some element $k\in M_{24}$,
 $(g,h)\sim (k^{-1}gk,k^{-1}hk)$,  or related by a modular transformation
\be 
(g,h)\sim (g^ah^c,g^bh^d)\ ,\qquad \begin{pmatrix}
a & b\\ c & d
\end{pmatrix}\in SL(2,\ZZ)\ .
\ee 
It follows that it is sufficient to determine just $55$ twisted twining genera. Of these, $21$ can be chosen to be of the form 
$\phi_{e,h}$ and therefore correspond to the twining genera computed in
 \cite{Cheng:2010pq,Gaberdiel:2010ch,Gaberdiel:2010ca,Eguchi:2010fg}. As 
for the remaining $34$ `genuinely twisted' genera, $28$ of them must vanish due to one of the obstructions 
described in section~\ref{s:obst}. The remaining six twisted twining genera can be computed as discussed 
in the previous subsection and have the form
\begin{align*}
&\phi_{\rm 2B,8A}=2\frac{\eta(2\tau)^2}{\eta(\tau)^4}\vartheta_1(\tau,z)^2\ , & & \phi_{\rm 2B,4A}=4\frac{\eta(2\tau)^2}{\eta(\tau)^4}\vartheta_1(\tau,z)^2\ , \\ 
&\phi_{\rm 4B,4A_1}=2\sqrt{2}\frac{\eta(2\tau)^2}{\eta(\tau)^4}\vartheta_1(\tau,z)^2\ ,    && \phi_{\rm 4B,4A_2}=2\sqrt{2}\frac{\eta(2\tau)^2}{\eta(\tau)^4}\vartheta_1(\tau,z)^2\ ,\\ 
 &\phi_{\rm 3A,3B}=0\ ,    &&\phi_{\rm 3A,3A}= 0\ ,
\end{align*} where the subscripts denote the conjugacy classes of the elements $g,h$ (see \cite{Gaberdiel:2012gf} for more details).

\bigskip

Once all the twisted twining genera satisfying \eqref{modcond} and \eqref{conj1} are known, one has to verify 
that they admit a decomposition of the form \eqref{eqn:decomp}. More precisely, one has to show that, for each 
$g\in M_{24}$, there exist projective representations $\rho_{g,r}$ of the centraliser $C_{M_{24}}(g)$ that match 
with \eqref{eqn:decomp}. Furthermore, the projective equivalence class of these representations must be the one 
determined by the $3$-cocycle $\alpha$.

In \cite{Gaberdiel:2012gf}, the first 500 such representations  were computed for each twisted sector 
(see the ancillary files of the arXiv version of the paper), and were shown to satisfy these properties. 
The only virtual representations that were found correspond 
to the BPS states in the untwisted ($g=e$) sector that appeared already in the original Mathieu moonshine 
(see \eqref{BPSreps}). Using the methods of \cite{GannonMathieu}, it should be possible to prove the existence 
of the representations $\rho_{g,r}$ for all $r$, and to confirm that there are indeed no 
virtual representations beyond  the ones in \eqref{BPSreps}. In any case, these results already provide very convincing 
evidence in favour of generalised Mathieu moonshine.

\vspace{.5cm}
\section{Conclusions}

In this short note we have reviewed the construction of the twisted twining genera for Mathieu moonshine. 
As we have explained, the twisted twining genera we have constructed behave very analogously to the 
twisted twining characters of holomorphic orbifolds; in particular, the various transformation properties of both are controlled by an 
element in $H^3(G,U(1))$. We regard this as convincing evidence for the idea that some (superconformal) VOA should
underlie and explain Mathieu moonshine. However, as we have also mentioned, this VOA cannot just be 
a sigma-model on K3, and it must have some unusual features in order to evade the arguments at the end of 
section~3.1. Understanding the structure of this VOA is, in our opinion, the central open problem in 
elucidating Mathieu moonshine.

\section*{Acknowledgments}

We thank Miranda Cheng, Mathieu Dutour, Graham Ellis, Terry Gannon, Jeff Harvey,  Stefan Hohenegger, Axel Kleinschmidt, Geoffrey Mason,
Ashoke Sen  and Don Zagier for useful conversations and correspondences. We also thank Henrik Ronnellenfitsch for the collaboration 
\cite{Gaberdiel:2012gf} on which this review is largely based.

\label{concl}

\end{document}